\documentclass[11pt,twoside]{article}

\usepackage{asp2014}

\aspSuppressVolSlug
\resetcounters
\begin{document}
\title{GDL today: Reaching a viable alternative to IDL.}
\author{Gilles Duvert$^{1}$, Alain Coulais$^{2}$ and Mark SChellens$^{3}$
\affil{$^1$Univ. Grenoble Alpes, IPAG, F-38000 Grenoble, France}
\affil{$^2$LERMA, CNRS and Observatoire de Paris, Paris, France}
}

\paperauthor{Gilles Duvert}{Gilles.Duvert@univ-grenoble-alpes.fr}{0000-0001-8769-3660}{Univ. Grenoble Alpes and CNRS}{IPAG}{Grenoble}{Isere}{38000}{France}
\paperauthor{Alain Coulais}{Alain.Coulais@obspm.fr}{}{Observatoire de Paris and CNRS}{LERMA}{Paris}{Paris}{75014}{France}

\begin{abstract}
On behalf of the GDL team and its project leader Marc Schellens, we report the progresses made by GDL, the free clone of the proprietary IDL software, since ADASS XXIII. We argue that GDL can replace IDL for everyday use.
\end{abstract}


\section{GDL, a free clone of IDL}
The GNU Data Language, known as ``GDL'' is a free clone of IDL, the
``interactive data language'', todays a product of Harris$^\circledR$
Geospatial Solution. IDL, thanks to its (historical) attractiveness,
ease of use, numerous functions and quality of graphical outputs, has
been largely used during 4 decades by the scientific community,
noticeably in Geophysics and Astrophysics. In particular, many
astronomical instruments and space missions relied, and still rely, on
IDL for data processing pipelines and data display. GDL has been
initially developed (in 2002) in the hope it could provide in the long
run a free and open source access and maintainance to these data
processing pipelines. Soon it reached a state where it could be used
indifferently with IDL for small tasks and projects.

The status of GDL has been regularly presented at
ADASS~\citep{2014ASPC..485..331C,2012ASPC..461..615C,2010ASPC..434..187C}.
We report in this paper that the maturity of GDL, now in version
0.9.8, makes it useable by all researchers in replacement of IDL for
everyday use. It has already been reported in
\citet{2014ASPC..485..331C} that GDL could run a computation intensive
pipeline for the PLANCK space mission. Recently, the Jean-Marie
Mariotti Center used GDL to run the optical interferometry image
reconstruction program \texttt{WISARD}\citep{2005JOSAA..22.2348M,
  2008JOSAA..26..108M} behind its OImaging webservice.  A
non-negligible amount of publications acknowledge the use of GDL: 42
to date in the ``astronomy'' collection of NASA's ADS bibliographic
system.

\section{Why GDL in these Python days?}
It is easily conceived that GDL is useful as a free alternative to IDL
when it boils down to use a data reduction pipeline, instrument
interface, simple handy script given by a colleague, etc\ldots written
in IDL. This use does not mean one adopts GDL as a scripting language,
just that the pipeline, interface or script does execute well under
GDL and produce the intended results.

But IDL (hence, GDL) is a very powerful vectorized, numerical and
interactive language. It is very fast for all vectorial operations and has
excellent (read: no-nonsense, publication-grade) graphic
outputs. Better, IDL (hence, GDL) knows what a scientific datum is. It
will not choke on the presence of NaNs in a measurement series, and
will, for example, correctly evaluate the mean and variance of this
data, or plot the measurement points with adequate bounding boxes and
scaling. For serious and frequent data analysis, this is invaluable
and it is natural that research fields where data are complicated,
error prone and/or undersampled, such as Astrophysics, have a long
partnership with IDL. Now GDL brings an interesting subset of IDL
possibilities for free in these disciplines.
\section{Recent Improvements}
\subsection{SPEED}
Today GDL is FAST. It was already benefitting from the Eigen library
\citep{eigenweb} since version 0.9.4 (2014), providing huge speedup in
matrix multiplication and some other operations---see
\citet{2014ASPC..485..331C}.  Now several basic functions have been
rewritten for speed: \texttt{SMOOTH}, \texttt{TRIANGULATE} and
\texttt{TRIGRID}, \texttt{WHERE}, \texttt{MOMENT}, \texttt{MEDIAN},
\texttt{MEAN}, \texttt{CONVOL}, \texttt{POLY\_2D},
\texttt{ISHFT}\ldots using if available the most recent and fast
open-source code available. As a rather extreme example, GDL uses now
the constant-time median filtering code \texttt{ctmf} of
S. Perreault~\citep{ctmf} for \texttt{MEDIAN} filtering
of BYTE arrays (images). \texttt{ctmf} makes for a noticeable speed
gain over IDL for large median width: $0.25$s
(GDL) vs. $7$s
(IDL) for a $300\times300$
median filtering of a $3000\times3000$
image. All the GDL performances are not as striking of course, but
running the \texttt{time\_test4.pro} procedure of IDL shows a marginal
gain of speed in favor of GDL ($1.09$s vs. $1.29$s).

Also, the speed of graphic outputs has been largely improved, and this
is very obvious on ``remote'' displays (local graphic windows fed by a
GDL running remotely---through an \texttt{ssh} session for
example). To do this, we had to rewrite the most time-consuming
functions of the \texttt{plplot} library GDL uses. We however seek to abandon
this library in the long run as we encounter more incompatibilities at
each \texttt{plplot} release.
\articlefigure[width=\textwidth]{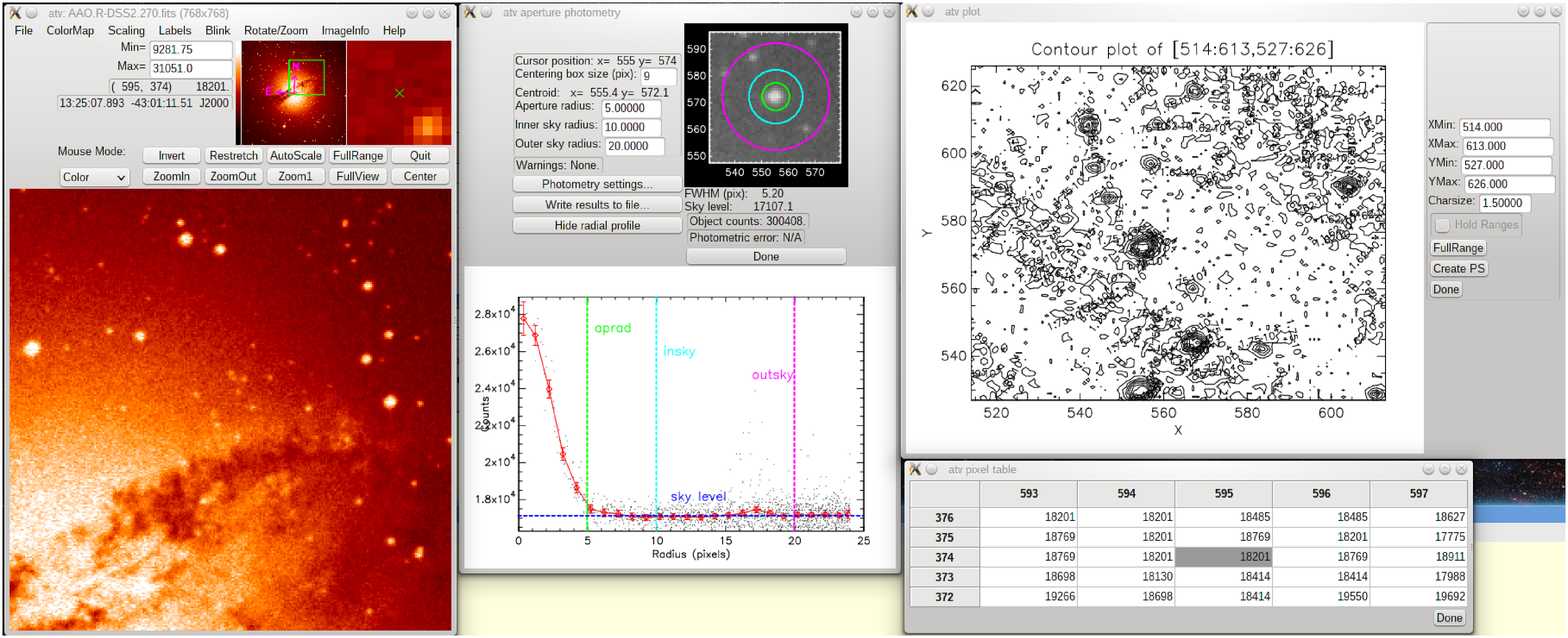}{fig:widgets}{Example of GDL widgets: the interactive display tool ATV
  showing the main window displaying an astronomical image, the aperture photometry panel, a region contour panel and
  the interactive pixel table panel.}
\subsection{WIDGETS}
The absence of widgets in GDL was becoming unsufferable as many
instrumental pipelines rely on them.  GDL has now (since 0.9.7) a full
set of WIDGETS using the wxWidgets library. Figure~\ref{fig:widgets}
shows 4 widgets of the ATV~\citep{2001ASPC..238..385B} astronomical
display tool. The development of these widgets was done differently to
previous GDL developments since we tried to reproduce all and every
feature of the base widgets objects of IDL, which is expressed in the
rich language options of the widget section of the IDL
documentation. This permitted GDL to run transparently and without
rewriting all the so-called ``compound'' widgets (higher level
meta-widgets created by an IDL procedure). See
section~\ref{sec:environment} for complementary information. We
encourage GDL contributors to follow this approach for further
developments.

Using wxWidgets for graphic widgets (\texttt{WIDGET\_DRAW}) had the
additional effect to add a wxWidgets graphics library to GDL. It is
now possible to use wxWidgets windows (as opposed to X11) by setting
the environment variable \texttt{GDL\_USE\_WX} to \texttt{YES}. Plots
using wxWindows benefit from antialiasing and system (non-Hershey)
fonts. It is expected that this will aid the porting to Microsoft's
``Windows'' operating system easier.
\subsection{Projections}
GDL supports now the full set of IDL projections in all plotting
functions.  However the \texttt{MAP\_SET} and \texttt{MAP\_PROJ\_INIT}
procedure that permit to set a projection have not been rewritten in a
license-free version, contrary to \texttt{MAP\_CONTINENTS} which is
part of GDL compiled code since 2010. This is not really an
impediment, as described in Sect~\ref{sec:environment}.
\subsection{Optimization}
Under this topic one finds the \texttt{AMOEBA}, \texttt{DFPMIN},
\texttt{POWELL} and \texttt{SIMPLEX} commands. These have been
implemented in version 0.9.8. \texttt{AMOEBA} uses the Simplex
algorithm of Nelder and Mead of the GSL, while
\texttt{DFPMIN} uses its version of the 
Broyden-Fletcher-Goldfarb-Shanno algorithm. \texttt{POWELL} uses
J. Burkard's C++ version of the PRAXIS method by R. Brent. Finally,
\texttt{SIMPLEX} is available if GDL is compiled with the GNU
\texttt{GLPK} library. All this is C++ compiled code and is expected
to outperform the equivalent IDL code based on Numerical Recipes
proprietary functions.
\section{Improving the GDL experience}\label{sec:environment}
\subsection{distribution issues}
As every open-source project that depends on external libraries, GDL
comes in different flavors depending on the computer type, software
distribution, etc\ldots Some libraries may be absent from a particular
distribution: currently the widget support in GDL is absent from
``homebrew'' on MacOSX, probably due to issues in building wxWidgets
on this platform. Similarly the Windows port lacks a number of
libraries.  Another difficulty is that GDL will be fast only when
built with optimisation on, which may not be the case on some
distributions. In all cases it is better, but more cumbersome, to build GDL
specifically for your computer, using CMake and the (limited)
documentation. See the README file and \url{http://aramis.obspm.fr/~coulais/IDL_et_GDL/minimum_script4gdl.html} for a minimal script to start with.

Reciprocally, some procedures written in IDL language, such as the
\texttt{idlcoyote} (\url{http://www.idlcoyote.com/documents/programs.php}) ones, test
the ``version'' of IDL, and for example refuse to operate if some
version-dependent capability is not present. GDL's versioning is
different, but the capability may be present. In this case, it is
possible to force GDL to report any IDL version number using the
\texttt{-\,-fakeversion} switch at GDL's invocation.
\subsection{Adding procedural libraries}
GDL comes with a very minimalist set of procedures (\texttt{.pro}
files) in its \texttt{!PATH}. It is recommended to add in the
\texttt{!PATH} the procedures of the \texttt{idlastro}
library available at
  \url{https://idlastro.gsfc.nasa.gov/}. Licensing reasons prevent us
to package GDL with many useful procedures, but these can be retrieved
independently, e.g., by googling. It is important to retrieve the CMSV
library of C. Markwardt available at
  \url{http://www.physics.wisc.edu/~craigm/idl/idl.html} to enable
the \texttt{SAVE} and \texttt{RESTORE} commands. Many, if not all,
procedures that come with the purchase of IDL can be used
transparently with GDL. Rewriting them for GDL, avoiding any licensing
problem, is a long task that calls for a serious help of new
contributors. In view of the possibility to find these procedures
elsewhere, this is not at the moment a very strong priority.
\subsection{Documentation}
The GDL documentation also is very minimal and would, as above,
benefit from new contributors to the project. Since GDL is intended to
be fully compatible with IDL, even for undocumented features and
behaviour, its implicit documentation is perforce the IDL one,
available on the net. The discussions lists on the GDL SourceForge pages are
also a useful location to find information.
\acknowledgements  The GDL team wishes to thank all users that filed bug reports as well as patches and suggestions using the SourceForge webpages of the project. We are particularly grateful to the packagers of GDL, who make it possible to have GDL without recompiling our enormous  list of software dependencies.
\bibliography{P9-108} 
\end{document}